\documentclass[reprint,amsmath,amssymb,aps]{revtex4}
\usepackage{graphicx}
\usepackage{dcolumn}
\usepackage{bm}
\usepackage{romannum}
\usepackage[normalem]{ulem}
\usepackage{setspace} 

\begin{document}
\title{Pseudogap and weak multifractality in disordered Mott charge-density-wave insulator}

\author{Jianhua Gao$^1$}
\author{Jae Whan Park$^1$}
\author{Kiseok Kim$^2$}
\author{Sun Kyu Song$^{1,2}$}
\author{Fangchu Chen$^{3,4}$}
\author{Xuan Luo$^3$}
\author{Yuping Sun$^{3,5,6}$}
\author{Han Woong Yeom$^{1,2}$}
\email{yeom@postech.ac.kr}
\affiliation{$^1$Center for Artificial Low Dimensional Electronic Systems, Institute for Basic Science (IBS),  77 Cheongam-Ro, Pohang 790-784, Korea}
\affiliation{$^2$Department of Physics, Pohang University of Science and Technology, Pohang 790-784, Korea}
\affiliation{$^3$Key Laboratory of Materials Physics, Institute of Solid State Physics, Chinese Academy of Sciences, Hefei 230031, People's Republic of China}
\affiliation{$^4$University of Science and Technology of China, Hefei, 230026, People's Republic of China}
\affiliation{$^5$High Magnetic Field Laboratory, Chinese Academy of Sciences, Hefei 230031, People's Republic of China}
\affiliation{$^6$Collaborative Innovation Centre of Advanced Microstructures, Nanjing University, Nanjing 210093, People's Republic of China}

\date {\today}

\begin{abstract}
The competition, coexistence and cooperation of various orders in low-dimensional materials like spin, charge, topological orders and charge-density-wave has been one of the most intriguing issues in condensed matter physics.
In particular, layered transition metal dichalcogenides provide an ideal platform for studying such an interplay with a notable case of 1\emph{T}-TaS$_{2}$ featuring Mott-insulating ground state, charge-density-wave, spin frustration and emerging superconductivity together.
We investigated local electronic states of Se-substituted 1\emph{T}-TaS$_{2}$ by scanning tunneling microscopy/spectroscopy (STM/STS), where superconductivity emerges from the unique Mott-CDW state.
Spatially resolved STS measurements reveal that an apparent V-shape pseudogap forms at the Fermi Level (E$_{F}$), with the origin of the electronic states splitting and transformation from the Mott states, and the CDW gaps are largely preserved. The formation of the pseudogap has little correlation to the variation of local Se concentration, but appears to be a global characteristics.
Furthermore, the correlation length of local density of states (LDOS) diverges at the Fermi energy and decays rapidly at high energies. The spatial correlation shows a power-law decay close to the Fermi energy. Our statistics analysis of the LDOS indicates that our system exhibits weak multifractal behavior of the wave functions.
These findings strongly support a correlated metallic state induced by disorder in our system, which provides an new insight into the novel mechanism of emerging superconductivity in the two-dimensional correlated electronic systems.
\end{abstract} \maketitle

In low-dimensional electronic systems, the interplay between electron correlation, superconductivity and charge density wave (CDW) or spin density wave has attracted great research interests.
Outstanding examples are the unconventional superconductivity in cuprate \cite{Science-344-608,NatMater-14-37} and pnictide \cite{NatCommun-7-13879,NatCommun-4-2874,NatPhys-9-220} materials.
Another notable case is the layered transition-metal dichalcogenides (TMDs), which provides an ideal platform for investigating the interplay between the various electronic orders \cite{NatMater-17-504,NatNanotech-13-483,NatCommun-7-11038, NatNanotech-11-339,NatPhys-12-139,Science-350-1353}.
Among many TMDs materials, 1\emph{T}-TaS$_{2}$ is an even more intriguing example, which exhibits a wide variety of CDW phases with different degree of incommensuration\cite{Science-243-1703,PRL-64-1150,SciAdv-1-1500606,NatCommun-7-11442,PRL-118-247401,NatPhys-11-328,PRL-107-177402,PRL-105-187401,Science-344-177}.
In particular, a metal-insulator transition occurs when the fully commensurate CDW phase develops and a Mott insulating state forms, with its unique Mott-CDW ground state where the electron correlation would play a crucial role leading probably to exotic quantum states.
Recently, it has been found that the Mott-CDW ground state can be transformed into various quasimetallic and superconducting states
by applying high pressure \cite{NatMater-7-960}, charge doping \cite{NatNanotech-10-270,APL-104-252601,PRL-109-176403,PRL-112-206402} or isovalent substitution of S with Se \cite{NatCommun-6-6091,PhysRevB-88-115145,PRL-91-256404, APL-102-192602}.
However, the microscopic mechanism for the emerging superconductivity has been elusive.
Especially, the general understanding of the superconducting order brought by suppressing commensurate or incommensurate CDW orders has not been established.
Various phenomenological ideas were proposed such as the superconductivity due to conductive domain wall networks \cite{NatMater-7-960}, the disordered Mott insulator \cite{PRL-112-206402} or the ordered superstructure \cite{NatCommun-6-6091}, which do not converge at all.
This problem is closely related to the possible unconventional superconductivity issue raised in the 2D limit of the layered transition metal dichalcogenides \cite{NatMater-17-504,NatNanotech-13-483,NatCommun-7-11038, NatNanotech-11-339,NatPhys-12-139,Science-350-1353}.
More fundamentally, this system provides an important and unique platform to investigate how entangled Mott and CDW gaps evolve upon the charge doping, domain walls and the random disorder.

In the present work, we investigate the local electronic structure of Se-substituted 1\emph{T}-TaS$_{2}$ (1\emph{T}-TaSSe with a S:Se ration close to 1:1 and a maximized superconducting temperature) by STM/STS probing simultaneously the atomic lattice, charge modulations and spectral functions with high spatial and energy resolution.
The ordered superstructure was directly observed, in which the CDW order and its energy gaps are largely preserved. On the other hand, the melting of the Mott gap into a pseudogap was identified clearly, indicating an insulator to metal transition occurred in 1\emph{T}-TaSSe.
Moreover, we found that this metallic behavior is not correlated to the local Se concentration but has a global characteristics.
The domain walls are found to be not important in the pseudogap formation.
Furthermore, we evidenced that the local density of states (LDOS) exhibits a diverging spatial correlation length at the Fermi energy and weak multifractal characteristics of the wave functions.
The global melting of the Mott gap into metallic states together with the weak multifractality of the wave functions strongly support the disordered Mott insulator scenario in the present case.
Based on a few theoretical proposals \cite{PRL-98, Ann-2010, PRL-12,PhysRevB-92-174526}, we claim that the disordered Mott insulator scenario provides a novel mechanism responsible for the emerging superconductivity.
The present work suggests it promising to discover an exotic quantum state around the superconducting quantum critical regime with strong disorder in charge and strong frustration in spin \cite{PNAS-2017}. More generally, this work suggests that the disorder can be another controllable parameter in 2D material systems, providing a planform to manipulate the correlated system to realize unique properties, which is more interesting and attractive for both physics and functionality.

\textbf{Results}

\textbf{Structural characterization of the Se substituted 1\emph{T}-TaS$_{2}$}

\begin{figure*}
\includegraphics[width=16.0 cm]{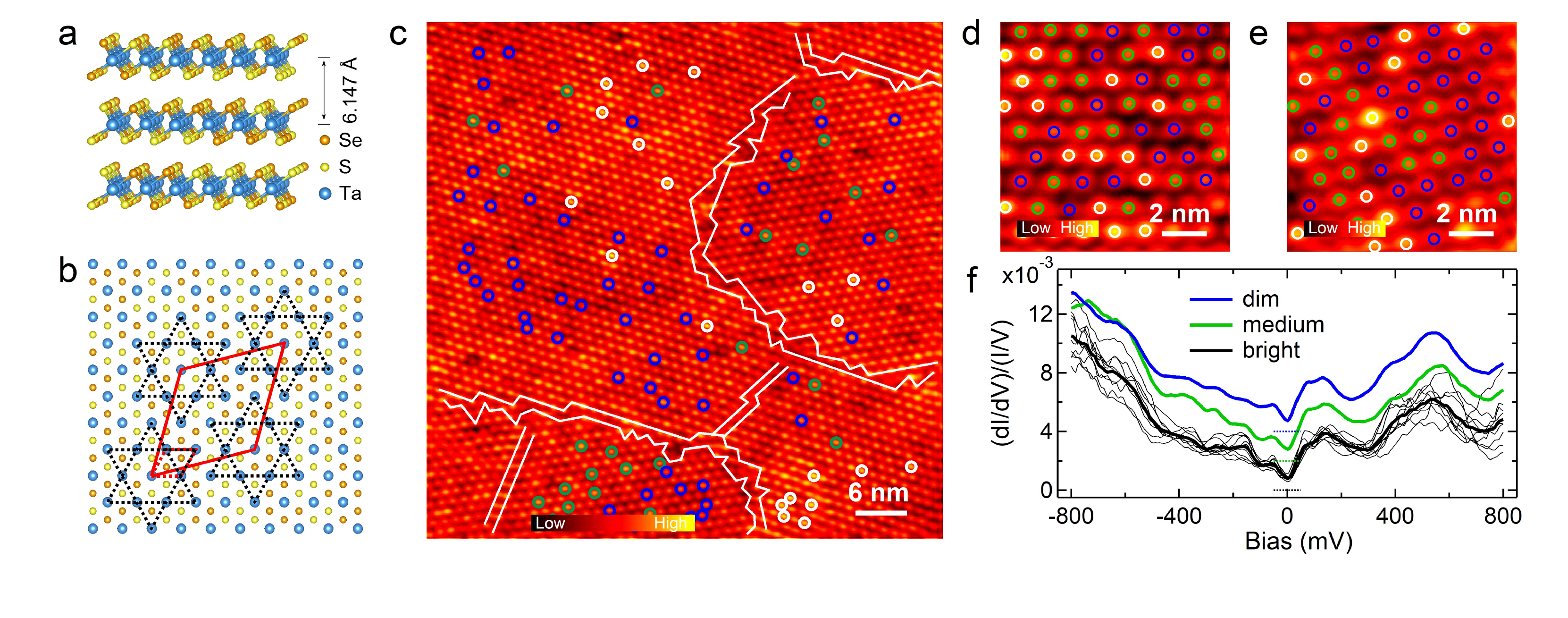}
\caption{\label{}Crystal structure and STS spectra of the Se-substituted 1\emph{T}-TaS$_{2}$ (1\emph{T}-TaSSe).
\textbf{a} The side-view schematics of the 1\emph{T}-TaSSe structure. Ta, S and Se atoms are depicted in blue, yellow and orange.
\textbf{b} The top-view where the pristine and CDW unit cells are indicated by dashed and solid lines in red color. The black dashed lines denote the Star of David pattern of the CDW clusters.
\textbf{c} STM topographic image of the 1\emph{T}-TaSSe surface acquired (\emph{V$_{s}$} = 800 mV and \emph{I$_{t}$} = 50 pA) at low temperature with a scanning area of 60$\times$60 nm$^{2}$.
It exhibits $\sqrt{13}\times\sqrt{13}$ R$13.9^{\circ}$ CDW phase, which splits into domains of two different orientations by a $30^{\circ}$ rotation between them.
The domain walls are marked by the white lines.
\textbf{d} and \textbf{e} display the enlarged (10$\times$10 nm$^{2}$) STM topographic images (\emph{V}$_{s}$ = 800 mV and \emph{I}$_{t}$ = 50 pA ).
The ordered CDW superstructure is preserved, but the brightness contrasts of the individual CDW maxima are obviously different.
The white, green and blue color circles correspond to bright, medium and dim contrasts in the images.
\textbf{f} \emph{dI/dV} spectra of the 1\emph{T}-TaSSe averaged over the CDW maxima.
The thick curves in black, green and blue color are averaged over CDW maxima in different groups as marked by white, green and blue circles in \textbf{c-e}.
The thin black curves are the spectra for individual CDW clusters of the corresponding group. }
\end{figure*}

Pristine 1\emph{T}-TaS$_{2}$ has a layered structure with a triangular lattice of Ta atoms sandwiched by layers of S atoms in an octahedral coordination \cite{NatCommun-7-10453,NatCommun-7-10956}.
At low temperature it develops a long-range CDW order where thirteen Ta atoms form a David-star CDW cluster as shown in Figs. 1\textbf{a} and 1\textbf{b}.
Se atoms are known to substitute randomly S atoms with the David-star CDW structure intact \cite{APL-102-192602}.
Each CDW cluster would have a different number of S or Se atoms.
Fig. 1\textbf{c} displays a topographic image of 1\emph{T}-TaSSe on a 60$\times$60 nm$^{2}$ area.
The image exhibits a periodic intensity variation, in which the bright contrast corresponds to the CDW maxima of each David star cluster.
The overall $\sqrt{13}\times\sqrt{13}$ CDW order can be identified clearly (Supplementary Fig. \textbf{S1} and \textbf{S2}) and the basic feature of the STM image is the same as 1\emph{T}-TaS$_{2}$.
The periodicity of the CDW superstructure can be extracted from the line profile (Supplementary Fig. \textbf{S1}), which is about 1.238 $\pm$ 0.009 nm larger than that of pristine 1\emph{T}-TaS$_{2}$ (1.21 nm) due to the larger atomic radius of Se. Furthermore, another obvious difference from 1\emph{T}-TaS$_{2}$ is the irregular domain boundaries between the translated or rotated CDW domains, as indicated by white lines in Fig. 1\textbf{c}.
The domain size varies between roughly 20 and 50 nm.
Moreover, a closer inspection in Figs. 1\textbf{d} and 1\textbf{e} reveals that the brightness of the CDW maxima are different from a unitcell to the other.
This is thought to be due to the difference in the local Se concentration, as detailed in a recent STM work \cite{PRX-7-041054}.
We quantify the topographic height (contrast) difference in the STM images with various different bias and classify the unitcells into four different groups (Supplementary Fig. \textbf{S3}).
Three of them are indicated in Figs. 1\textbf{c}, 1\textbf{d} and 1\textbf{e} by different color circles. The white, green and blue color circles correspond to bright, medium and dim contrasts, respectively.

\textbf{Spectroscopic characterization of the Se substituted 1\emph{T}-TaS$_{2}$}

In order to investigate detailed electronic structures, spatially resolved STS measurements were performed on various CDW unitcells with different topographic contrast as shown in Figs. 1\textbf{f} and Supplementary Fig. \textbf{S4}.
The spectra from the CDW clusters within each group with similar brightness contrasts show a rather large variation between unitcells (Fig. 1\textbf{f} and Supplementary Fig. \textbf{S4}), which is not observed in pristine 1\emph{T}-TaS$_{2}$ \cite{PhysRevB-92-085132}.
However, the averaged spectra for CDW clusters with similar contrast are compared with those of different groups, as shown in Fig. 1\textbf{f} (also Supplementary Fig. \textbf{S4}).
Obviously, the difference between groups is not larger than that between individual CDW clusters within a given group, which indicates that the features of the spectra are not correlated with the topographic contrasts.
The topographic contrasts, that is, the corrugation of S-Se layer, was suggested to represent the Se concentration of each CDW cluster. Therefore, the local electronic structure is not correlated with the local Se concentration.
This result is contradictory to the very recent work, which reported that the local electronic states of the individual clusters are highly dependent on their topographic contrasts with locally defined band gaps \cite{PRX-7-041054}.
However, no systematic variation of the spectra feature depending on the local topographic contrasts was observed in our measurements.
This point will be discussed further below.

Instead, the spectra averaged over CDW clusters are rather well established in spite of the variation in the corrugation of the S-Se layer, especially for the spectral features near the Fermi energy.
Figs. 2\textbf{a} and 2\textbf{b} compare the \emph{dI/dV} spectra averaged over many CDW maxima and minima, respectively.
This can also be corroborated visually in the measured LDOS maps over ten neighboring CDW unitcells in Fig. 2 \textbf{c-h} (also Supplementary Fig. \textbf{S5}).
All these LDOS maps for major spectral features indicate the CDW pattern clearly.
It is thus obvious that the local variation of the Se concentration does not override the periodic spatial modulation of the major spectral features.

\textbf{Pseudogap formation of the Se substituted 1\emph{T}-TaS$_{2}$}

However, at the same time, the overall spectral change induced by the Se substitution is very prominent in the averaged spectra.
The spectra of the 1\emph{T}-TaS$_{2}$ feature the multiple gap structure, notably the Mott gap with upper and lower Hubbard band (UHB and LHB) states at +230 and -130 mV and the CDW gap edges at +420 mV and -370 mV \cite{PhysRevB-92-085132}.
In contrast, 1\emph{T}-TaSSe exhibits a rather enhanced CDW gap with edges at +530 mV and -400 mV.
The separation between UHB state and the upper CDW gap edge is noticeably widened, which is also the common feature of DFT calculations for Se-substituted CDW clusters (Supplementary Figs. \textbf{S6} and \textbf{S7}).
More importantly, the Mott states exhibit strong spectral modulation.
Both UHB and LHB states split into mainly two structures: the major ones appear at slightly lower energy by $\sim$30 meV from those on 1\emph{T}-TaS$_{2}$ with substantially reduced weights and extra states appear at energies closer to the Fermi level (marked by arrows in Fig. 2\textbf{a}).
Note that there exists finite density of states at the Fermi energy tailed out of the newly formed spectral features within the Mott gap.
It is obvious that the spectral weights of the UHB and LHB states are transformed to these in-gap states, which forms a V-shape energy gap of 130 mV and metallic DOS at Fermi energy.
This is a clear spectroscopic indication of a Mott-insulator-to-metal transition through the formation of a pseudogap.

\begin{figure}
\includegraphics[width=8.0 cm]{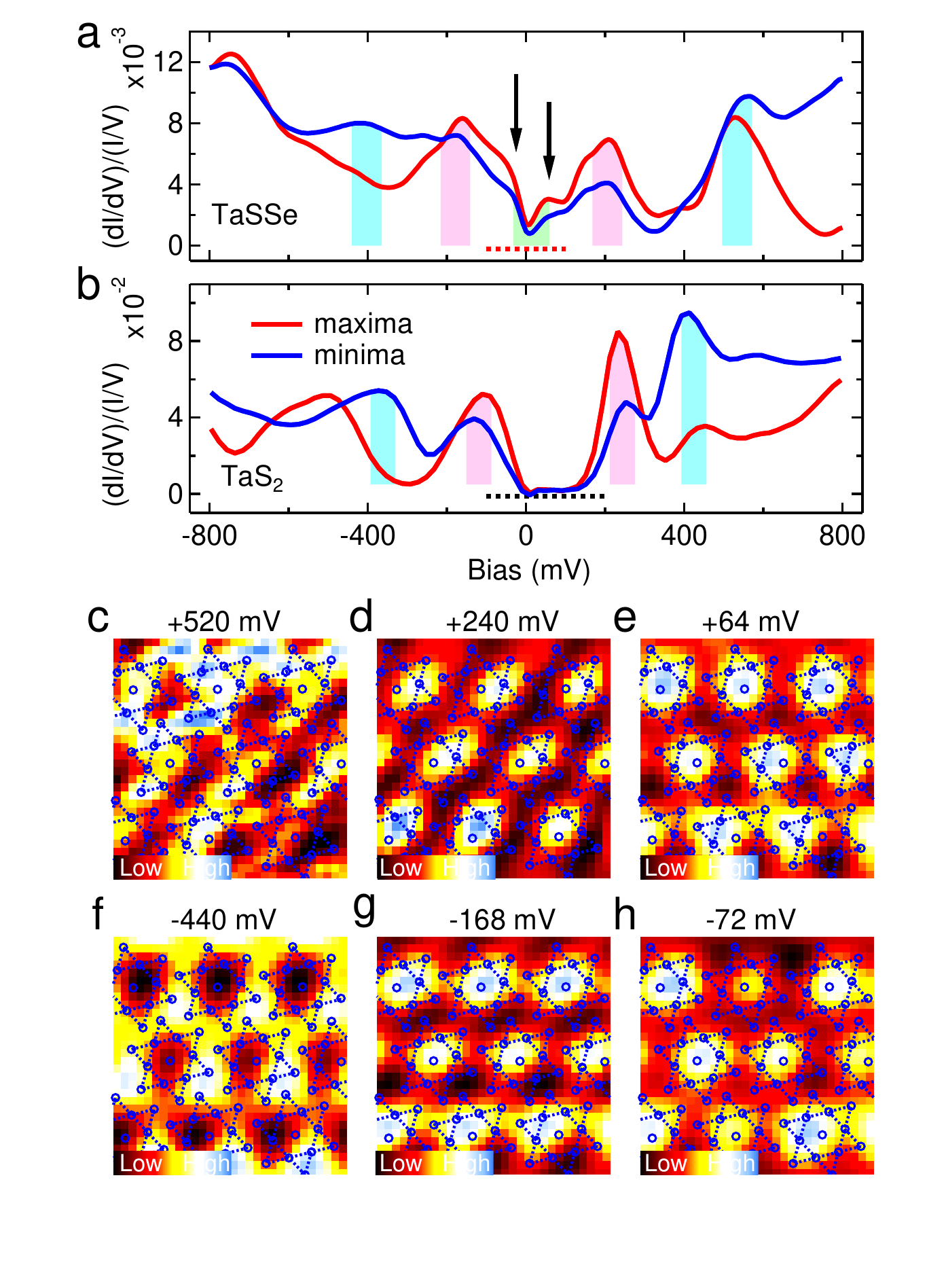}
\caption{\label{} Averaged STS spectra and LDOS maps of the 1\emph{T}-TaSSe.
\textbf{a} and \textbf{b} show the normalized STS spectra of the 1\emph{T}-TaSSe and 1\emph{T}-TaS$_{2}$, respectively.
The red (blue) curves correspond to the spectra from the CDW maxima (minima).
Blue and pink bars denote the CDW and Mott states. The green bar and arrows indicate the pseudogap state of the 1\emph{T}-TaSSe.
\textbf{c}-\textbf{h} illustrate LDOS maps of the 1\emph{T}-TaSSe over an area of 3.6$\times$3.6 nm$^{2}$ at representative energies acquired from the grid \emph{dI/dV} measurements (\emph{V}$_{s}$ = $\pm$ 800 mV and \emph{I}$_{t}$ = 900 pA).}
\end{figure}

The origin of the major spectral features can further be checked by their spatial characteristics revealed in the LDOS maps (Fig. 2 \textbf{c-h}).
The LDOS maps at energies of +520 and -440 mV have strong enhancements at CDW minima, which is consistent with the spatial variation of CDW gap edges in the pristine 1\emph{T}-TaS$_{2}$ \cite{PhysRevB-92-085132}.
In clear contrast, the states at +240 and -168 mV are enhanced at CDW maxima.
This is consistent with the spatial variation of UHB and LHB states in the pristine sample\cite{PhysRevB-92-085132} as they are localized on central Ta atoms of David-star CDW clusters.
Very importantly, the newly formed in-gap states at +64 and -72 mV are also concentrated on the CDW maxima regions.
This spatial characteristics confirms the idea that the pseudogap spectral weights are transferred from UHB and LHB states.

\begin{figure*}
\includegraphics[width=16.0 cm]{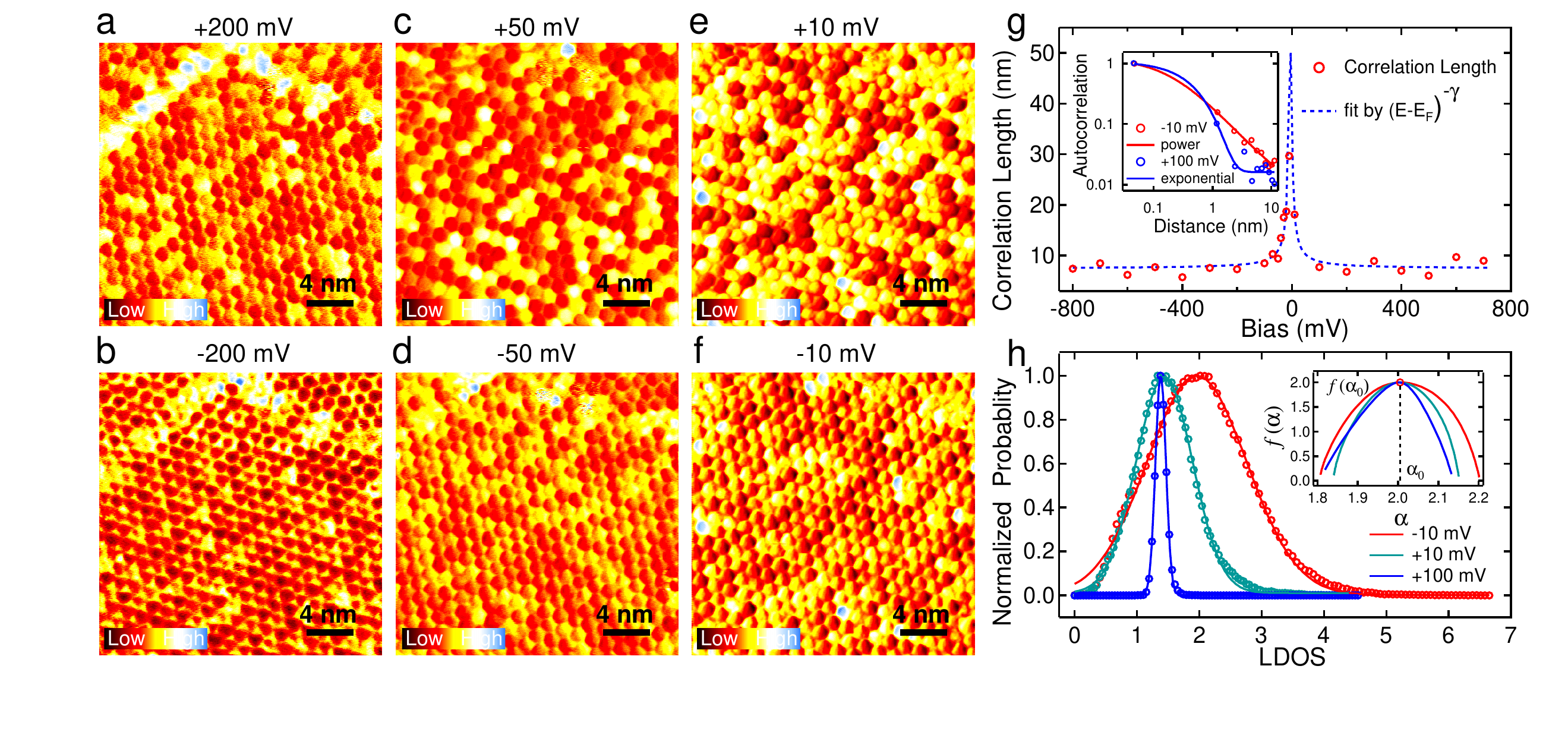}
\caption{\label{} The spatial correlation and multifractal behavior of the electronic states.
The LDOS maps were taken with the scanning area of 24$\times$24 nm$^{2}$ at \textbf{a} \emph{V}$_{s}$ = +200 mV, \emph{I}$_{t}$ = 50 pA; \textbf{b} \emph{V}$_{s}$ = -200 mV, \emph{I}$_{t}$ = 50 pA; \textbf{c} \emph{V}$_{s}$ = +50 mV, \emph{I}$_{t}$ = 50 pA; \textbf{d} \emph{V}$_{s}$ = -50 mV, \emph{I}$_{t}$ = 50 pA; \textbf{e} \emph{V}$_{s}$ = +10 mV, \emph{I}$_{t}$ = 50 pA; \textbf{f} \emph{V}$_{s}$ = -10 mV, \emph{I}$_{t}$ = 50 pA.
\textbf{g} The correlation length extracted from the LDOS maps as a function of the energies.
The dashed line in blue color is the fitted curve by following (E-E$_{F}$)$^{-\gamma}$, with $ \gamma = 1.08\pm0.01 $.
The inset shows the decay of the spatial correlation at two representative energies.
The red and blue color lines correspond to the energy at - 10 mV and + 100 mV, respevtively.
\textbf{h} The normalized histogram of the LDOS maps at -10 mV, +10 mV and +100 mV, as denoted by red, green and blue color circles.
The solid lines correspond to the Gaussian fittings.
The inset displays multifractal spectra \emph{f}($\alpha$) calculated from the same LDOS maps at -10 mV, +10 mV and +100 mV.
The multifractal spectra \emph{f}($\alpha$) exhibit the parabolic shape with the maximum position locating at $ d + \epsilon $ (where $ d=2.0 $ and $ \epsilon \ll 1$), indicating the weak-multifractality behavior of the electronic states.}
\end{figure*}

\textbf{Spatial correlation of the electronic states and weak multifractality in the Se substituted 1\emph{T}-TaS$_{2}$}

After establishing the global spectral function of 1\emph{T}-TaSSe, we turn our interest into the spatial fluctuation.
The \emph{dI/dV} maps were acquired over an area of 24 nm$\times$ 24 nm at representative energies, which are shown in Figs. 3\textbf{a}-\textbf{f}.
The LDOS enhancement on a domain wall \cite{NatCommun-7-10453} is clear at an empty state of + 200 meV while it does not leave any footprint nearer to the Fermi level.
Therefore, the role of domain walls in the metallic property and the emerging superconductivity is excluded in 1\emph{T}-TaSSe.
On the other hand, as mentioned above, the LDOS (and STS spectra) exhibits the unitcell-by-unitcell fluctuation.
Moreover, these spatial fluctuation patterns are not consistent at different energies and with the topographic contrast variaions (Supplementary Fig. \textbf{S3}). These results indicate that the electronic states are not spatially related to the local Se concentration and are of random character in both energy and space.
The amplitude of fluctuations is 20-40 \% of the average LDOS values (Supplementary Fig. \textbf{S8}), with different spatial variations in the LDOS maps at different energies. However, the histogram of the LDOS are similar in following Gaussian distribution (Figs. 3\textbf{h}).
This is a characteristic signature of disordered extended electronic states, indicating that these states can be present over the entire system, which are confirmed by STM spectra previously.
Furthermore, the energy and spatial dependence of the fluctuation can further be quantified by extracting the correlation length in the angle-averaged autocorrelation function of the LDOS maps \cite{Science-327-665}.
The extracted correlation lengths as a function of the energies are shown in Fig. 3\textbf{g}, which diverges at the Fermi energy and are suppressed away from E$_{F}$. The correlation lengths as a function of the energies follow $ (E-E_{F})^{-\gamma} $, with $ \gamma = 1.08\pm0.01 $. 
The spatial correlation is established only very close to the Fermi level and decays rapidly at higher energies.
The inset of Fig. 3\textbf{g} shows the distance dependence of the correlation (spatial correlation) at different energies, which follows power-law decay close to the Fermi energy (at -10mV) and fall off exponentially away from the Fermi energy (at +100mV). 
These observations together with the divergence of the correlation lengths are indication of the present system close to the critical regime associated with metal-insulator transition, which usually lead to the electronic states exhibiting multifractal behavior.

The multifractality is related to the scale invariant nature of critical wave functions near Anderson transition.
The nature of the critical wave functions can be further analyzed by their multifractal patterns through their singularity spectrum \emph{f}($\alpha$)\cite{Science-327-665}, which describes all the fractal dimensions embedded in a spatial pattern and its probability distribution.
The analysis of the multifractal spectra \emph{f}($\alpha$) at different energies are shown in the inset of Fig. 3\textbf{h} (also Supplementary Fig. \textbf{S9} ).
In all the cases, the multifractal spectra \emph{f}($\alpha$) exhibit the parabolic shape with the maximum position at $ \alpha_{0}= d + \epsilon $ (where $ d= 2.0 $ is the dimensional of the system and $ \epsilon =0.008, 0.005, 0.003$ at different energies, which satisfy $ \epsilon \ll 1$) demonstrating the weak-multifractality behavior, as expected for extended electronic states in two-dimensional systems \cite{PRL-98, Ann-2010, PRL-12}.
To the best of our knowledge, the present system is the first one exhibiting the weak multifractality with the strong interaction.
These behaviors are consistent signatures of disordered electronic states being close to the critical regime of an Anderson metal-insulator transition \cite{RMP-57-287,RMP-80-1355,RMP-66-261}.

\textbf{Discussion}

In summary, we investigated the Mott insulator to metallic state of 1\emph{T}-TaS$_{2}$ upon isovalent Se substitution, in which the CDW gaps are largely preserved and the formation of the pseudogap has no correlation with the local Se concentration. We interpret the statistical analysis of the electronic state correlation in terms of multifractal wave function, which originates from the disorder.
Our present findings are contradictory to the recent results\cite{PRX-7-041054} of the Mott gap collapse by shifting the Hubbard states through the Se substitition, which induces a charge transfer insulator.
The authors claimed that the CDW clusters exhibit different local band gaps with very few metallic ones scattered\cite{PRX-7-041054}.
We should note that such few isolated metallic clusters are not consistent at all with the emergence of superconductivity.
Moreover, their results are not consistent with the ARPES study reporting the formation of an electron pocket\cite{PhysRevB-88-115145}, which is possibly due to the deviation from the critical amount of disorder for the formation of a global pseudogap.
Our own DFT calculations confirm this result but a very similar spectral change can also be induced by different S-Se configurations with the same Se concentration (Supplementary Figs. \textbf{S6} and \textbf{S7}).
However, such a spectral behavior is not reproduced at all in the present experiments.
We note that DFT calculations deal only with fully regular periodic structures of a given concentration and a fixed S-Se configuration, while the reality is a largely disordered concentration and configuration over CDW clusters prohibiting any periodicity in principle.
Futhermore, our present results is also not consistent with the connection of the ordered superstructure found in the TEM study to the superconductivity \cite{NatCommun-6-6091}.
Such an extra ordering was not found in both the present work and the recent STM study\cite{PRX-7-041054}.
More importantly, the CDW gaps are well preserved or even enhanced in our experiments, which indicates that the collapse of the Mott state is not controlled by the CDW order.
Therefore, our findings obviously suggests that the emerging superconductivity of the present system does not follow the conventional wisdom of competing orders, calling for an exotic mechanism.

On the other hand, the present result is consistent with the ARPES observation of the pseudogap states in disordered 1\emph{T}-TaS$_2$ by random Cu intercalants \cite{PRL-112-206402}, in which the pseudogap spectral function expected theoretically is very similar to what we observed here.
The disorder effect on a Mott insulator has been extensively investigated, which leads generally to correlated metallic states with pseudogap features \cite{NatPhys-13-21}.
In our work, we evidenced that the LDOS exhibit a diverging spatial correlation at the Fermi level and weak multifractal behavior, which makes the disorder origin more clearly.
In parallel, the general wisdom, so called Anderson theorem, proposed that the superconductivity is largely immune to disorder \cite{Anderson}.
However, only very recently, it was theoretically suggested that the multifractality of the wave functions near the localization transition can enhance superconductivity  \cite{PRL-98, Ann-2010, PRL-12}.
The disorder-induced superconductivity was also discussed in 2\emph{H}-TaSe$_2$ and 2\emph{H}-NbSe$_2$ experimentally. \cite{NPJQM-2-11,NatPhys-12-92,arXiv-1810}
This new insight provides the microscopic mechanism of the emerging superconductivity and we think it could potentially be important for the intriguing superconductivity in our present system.
Moreover, the tunability of the disorder by the Se concentration makes the present system even more attractive. 
A disordered correlated metallic state established here provides an interesting platform to find new many body quantum states such as those suggested in the presence of dilute magnetic impurities \cite{PRL-99, PRB-85}.

\textbf{Methods}

\textbf{Experimental details}

The high-quality single crystals of 1\emph{T}-TaSSe (S/Se ratio being close to one-to-one) were synthesized by chemical vapor transport method with iodine as the transport agent.
Details of the sample preparation were described elsewhere \cite{APL-102-192602}.
The crystals were cleaved \emph{in situ} at room temperature in an ultrahigh vacuum chamber equipped with a commercial cryogenic STM (SPECS, Germany). STM experiments were performed by mechanically sharpened Pt-Ir wires at 4.3 K.
The topographic images were taken in constant current mode with bias voltage \emph{V$_{s}$} applied to the sample,
and the \emph{dI/dV} spectra were acquired by using the lock-in technique with a voltage modulation of 5 mV and a frequency modulation of 1.0 kHz.

\textbf{Calculation details}

DFT calculations were performed by using the Vienna $ab$ $initio$ simulation package \cite{PRB-54-11169} within the Perdew-Burke-Ernzerhof generalized gradient approximation \cite{PRL-77-3865} and the projector augmented wave method \cite{PRB-50-17953}.
The single-layer 1$T$-TaS$_{2-x}$Se$_{x}$ was modeled with a vacuum spacing of about 13.6 {\AA}.
We used a plane-wave basis set of 260 eV and a 8$\times$8$\times$1 $k$-point mesh for the  $\sqrt{13}\times\sqrt{13}$ unit cell and atoms were relaxed until the residual force components were within 0.01 eV/{\AA}.
To more accurately represent the electronic correlation we used an on-site Coulomb energy of 1.5 eV for Ta 5$d$ orbitals, which reproduces the experimental Mott gap size of 1$T$-TaS$_{2}$.

\textbf{Acknowledgements}

This work was supported by Institute for Basic Science (Grant No. IBS-R014-D1). H.W.Y. appreciates the enlightening discussion with Changwon Park. F.C.C., X.L. and Y.P.S. thank the support of the National Key Research and Development Program under contracts 2016YFA0300404 and the National Nature Science Foundation of China under contracts 11674326, 11874357 and the Joint Funds of the National Natural Science Foundation of China and the Chinese Academy of Sciences Large-Scale Scientific Facility under contract U1832141.

\textbf{Author contributions}
H.W.Y. conceived the research idea and plan.
G.J.H. performed the STM experiments through discussion with H.W.Y.
C.F.C. grew the single crystals under the supervision of X.L. and Y.P.S.
J.W.P. performed the DFT calculations.
K.K provided theoretical ideas.
G.J.H and H.W.Y. prepared the manuscript with the comments of all other authors.


\begin{thebibliography}{99}

\bibitem {Science-344-608} He, Y. \emph{et al}. Fermi surface and pseudogap evolution in a cuprate
superconductor. Science 344, 608-611 (2014).

\bibitem {NatMater-14-37} Hashimoto, M. \emph{et al}. Direct spectroscopic evidence for phase competition between the pseudogap and superconductivity in Bi$_{2}$Sr$_{2}$CaCu$_{2}$O$_{8+\delta}$, Nat. Mater. 14, 37-42 (2015).

\bibitem {NatCommun-7-13879} Song, Y. \emph{et al}. A Mott insulator continuously connected to iron pnictide superconductors, Nat. Commun. 7, 13879 (2016).

\bibitem {NatCommun-4-2874} Wang, M. \emph{et al}. Doping dependence of spin excitations and its correlations with high-temperature superconductivity in iron pnictides, Nat. Commun. 4, 2874 (2013).

\bibitem {NatPhys-9-220} Allan, M. P. \emph{et al}. Anisotropic impurity states, quasiparticle scattering and nematic transport in underdoped Ca(Fe$_{1-x}$Co$_{x}$)$_{2}$As$_{2}$, Nat. Phys. 9, 220-224 (2013).

\bibitem {NatMater-17-504} Sohn, E. \emph{et al}. An unusual continuous paramagnetic-limited superconducting phase transition in 2D NbSe$_{2}$, Nat. Mater. 17, 504-508 (2018).

\bibitem {NatNanotech-13-483} Costanzo, D. \emph{et al}. Tunneling spectroscopy of gate-induced superconductivity in MoS$_{2}$, Nat. Nanotech. 13, 483-488 (2018).

\bibitem {NatCommun-7-11038} Qi, Y. P. \emph{et al}. Superconductivity in Weyl semimetal candidate MoTe$_{2}$, Nat. Commun. 7, 11038 (2016).

\bibitem {NatNanotech-11-339} Costanzo, D. \emph{et al}. Gate-induced superconductivity in atomically thin MoS$_{2}$ crystals, Nat. Nanotech. 11, 339-344 (2016).

\bibitem {NatPhys-12-139} Xi, X. X. \emph{et al}. Evidence of Ising pairing in superconducting NbSe$_{2}$ atomic layers, Nat. Phys. 12, 139-143 (2016).

\bibitem {Science-350-1353} Lu, J. M. \emph{et al}. Evidence for two-dimensional Ising superconductivity in gated MoS$_{2}$, Science, 350, 1353-1357 (2015).



\bibitem {Science-243-1703} Wu, X. L. \emph{et al}. Hexagonal domain-like charge density wave phase of TaS$_{2}$ determined by scanning tunneling microscopy Science, 243, 1703-1705 (1989).
\bibitem {PRL-64-1150} Wu, X. L. \emph{et al}. Direct observation of growth and melting of the hexagonal-domain charge-density-wave phase in 1\emph{T}-TaS$_{2}$ by
scanning tunneling microscopy. Phys. Rev. Lett. 64, 1150-1154 (1990).

\bibitem {SciAdv-1-1500606} Yoshida, M. \emph{et al}. Memristive phase switching in two-dimensional 1T-TaS$_{2}$ crystals, Science Advances, 1, 1500606 (2015).

\bibitem {NatCommun-7-11442} Vaskivskyi, I. \emph{et al}. Fast electronic resistance switching involving hidden charge density wave states, Nat. Commun. 7, 11442 (2016).

\bibitem {PRL-118-247401} Laulh\'{e}, C. \emph{et al}. Ultrafast formation of a charge density wave state in 1\emph{T}-TaS$_{2}$: Observation at nanometer scales using time-resolved X-ray diffraction, Phys. Rev. Lett. 118, 247401 (2017). 

\bibitem {NatPhys-11-328} Ritschel, T. \emph{et al}. Orbital textures and charge density waves in transition metal dichalcogenides, Nat. Phys. 11, 328-331 (2015).

\bibitem {PRL-107-177402} Petersen, J. C. \emph{et al}. Clocking the melting transition of charge and lattice order in 1\emph{T}-TaS$_{2}$ with ultrafast extreme-ultraviolet angle-resolved photoemission spectroscopy, Phys. Rev. Lett. 107, 177402 (2011). 

\bibitem {PRL-105-187401} Hellmann, S. \emph{et al}. Ultrafast melting of a charge-density wave in the Mott insulator 1\emph{T}-TaS$_{2}$, Phys. Rev. Lett. 105, 187401 (2010).

\bibitem {Science-344-177} Stojchevska, L. \emph{et al}. Ultrafast switching to a stable hidden quantum state in an electronic crystal, Science 344, 177-180 (2014).

\bibitem {NatMater-7-960} Sipos, B. \emph{et al}. From Mott state to superconductivity in 1\emph{T}-TaS$_{2}$. Nat. Mater. 7, 960-965 (2008).

\bibitem {NatNanotech-10-270} Yu, Y. \emph{et al}. Gate-tunable phase transitions in thin flakes of 1\emph{T}-TaS$_{2}$. Nat. Nanotechnol. 10, 270-276 (2015).

\bibitem {PRL-109-176403} Ang, R. \emph{et al}. Real-space coexistence of the melted Mott state and superconductivity in Fe-substituted 1\emph{T}-TaS$_{2}$. Phys. Rev. Lett. 109, 176403 (2012).

\bibitem{APL-104-252601} Liu, Y. \emph{et al}. Coexistence of superconductivity and charge-density-wave domain in 1\emph{T}-Fe$_{x}$Ta$_{1-x}$SSe, Appl. Phys. Lett. 104, 252601 (2014).

\bibitem {PRL-112-206402} Lahoud, E. \emph{et al}. Emergence of a novel pseudogap metallic state in a disordered 2D Mott insulator, Phys. Rev. Lett. 112, 206402 (2014).

\bibitem {NatCommun-6-6091} Ang, R. \emph{et al}. Atomistic origin of an ordered superstructure induced superconductivity in layered chalcogenides. Nat. Commun. 6, 6091 (2015).

\bibitem{PhysRevB-88-115145} Ang, R. \emph{et al}. Superconductivity and bandwidth-controlled Mott metal-insulator transition in 1\emph{T}-TaS$_{2-x}$Se$_{x}$, Phys. Rev. B 88, 115145 (2013).

\bibitem {PRL-91-256404} Aiura, Y. \emph{et al}. Ta 5\emph{d} band symmetry of 1\emph{T}-TaS$_{1.2}$Se$_{0.8}$ in the commensurate charge-density-wave phase, Phys. Rev. Lett. 91, 256404 (2003).


\bibitem{APL-102-192602} Liu, Y. \emph{et al}. Superconductivity induced by Se-doping in layered charge-density-wave system 1\emph{T}-TaS$_{2-x}$Se$_{x}$, Appl. Phys. Lett. 102, 192602 (2013).

\bibitem{PRL-98} Feigel'man, M. V. \emph{et al}. Eigenfunction fractality and pseudogap state near the superconductor-insulator transition, Phys. Rev. Lett. 98, 027001 (2007).

\bibitem{Ann-2010} Feigel'man, M. V. \emph{et al}. Fractal superconductivity near localization threshold, Ann. Phys. 325, 1390 (2010).

\bibitem{PRL-12} Burmistrov, I. S. \emph{et al}. Enhancement of the critical temperature of superconductors by anderson localization, Phys. Rev. Lett. 98, 027001 (2012).

\bibitem{PhysRevB-92-174526} Mayoh, J. \emph{et al}. Global critical temperature in disordered superconductors with weak multifractality, Phys. Rev. B 92, 174526 (2015).

\bibitem{PNAS-2017} Lee, K. T. \emph{et al}. 1\emph{T}-TaS$_{2}$ as a quantum spin liquid, Proc. Nat'l. Acad. Sci. 114, 6996 (2017).
\bibitem {NatCommun-7-10453} Cho, D. \emph{et al}. Nanoscale manipulation of the Mott insulating state coupled to charge order in 1\emph{T}-TaS$_{2}$. Nat. Commun. 7, 10453 (2016).
\bibitem {NatCommun-7-10956} Ma, L. \emph{et al}. A metallic mosaic phase and the origin of Mott insulating state in 1\emph{T}-TaS$_{2}$. Nat. Commun. 7, 10956 (2016).

\bibitem {PRX-7-041054} Qiao, S. \emph{et al}. Mottness collapse in 1\emph{T}-TaS$_{2-x}$Se$_{x}$ transition-metal dichalcogenide: An interplay between localized and itinerant orbitals, Phys. Rev. X. 7, 041054 (2017).

\bibitem{PhysRevB-92-085132} Cho, D. \emph{et al}. Interplay of electron-electron and electron-phonon interactions in the low-temperature phase of 1\emph{T}-TaS$_{2}$, Phys. Rev. B 92, 085132 (2015).

\bibitem {Science-327-665} Richardella, A. \emph{et al}. Visualizing critical correlations near the metal-insulator transition in Ga$_{1-x}$Mn$_{x}$As, Science, 327, 665-669 (2010).

\bibitem {RMP-57-287}  Lee, P. A. \emph{et al}. Disordered electronic systems, Rev. Mod. Phys. 57, 287, (1985).
\bibitem {RMP-80-1355} Evers, F. \emph{et al}. Anderson transitions, Rev. Mod. Phys. 80, 1355 (2008).
\bibitem {RMP-66-261} Belitz, D. \emph{et al}. The Anderson-Mott transition. Rev. Mod. Phys. 66, 261 (1994).

\bibitem {NatPhys-13-21} Battisti, I. \emph{et al}. Universality of pseudogap and emergent order in lightly doped Mott insulators Nat. Phys. 13, 21-25 (2017).

\bibitem{Anderson} Anderson, P. W. Theory of dirty superconductors, J. Phys. Chem. Solids 11, 26 (1959).

\bibitem {NPJQM-2-11} Li, L. J. \emph{et al}. Superconducting order from disorder in 2\emph{H}-TaSe$_{2-x}$S$_{x}$, npj Quantum Materials 2: 11 (2017).

\bibitem {NatPhys-12-92} Ugeda, M. M. \emph{ et al}. Characterization of collective ground states in single-layer NbSe$_{2}$. Nat. Phys. 12, 92-97 (2016).

\bibitem {arXiv-1810} Verd\'{u}, C, R \emph{ et al}. Multifractal superconductivity in a two-dimensional transition metal dichalcogenide. arXiv: 1810.08222.

\bibitem{PRL-99} Zhuravlev, A. \emph{ et al}. Nonperturbative scaling theory of free magnetic moment phases in disordered metals, Phys. Rev. Lett. 77, 3865 (1996).

\bibitem{PRB-85} Kettemann, S. \emph{ et al}. Kondo-Anderson transitions, Phys. Rev. B 85, 115112 (2012).

\bibitem{PRB-54-11169} Kresse, G. \emph{ et al}. Efficient iterative schemes for $ab$ $initio$ total-energy calculations using a plane-wave basis set, Phys. Rev. B 54, 11169 (1996).

\bibitem{PRL-77-3865} Perdew, J. P. \emph{ et al}. Gerneralized gradient approximation made simple, Phys. Rev. Lett. 77, 3865 (1996).

\bibitem{PRB-50-17953} Blochl, P. E. Projector augmented-wave method, Phys. Rev. B 50, 17953 (1994).

\end{thebibliography}
\end{document}





\textbf{Supplementary Note 1: Morphology of the Se substituted 1\emph{T}-TaS$_{2}$}

Figure \textbf{S1} \textbf{a} shows a typical STM topographic image taken at 4.3 K within a single domain with the scanning area of 40$\times$40 nm$^{2}$ at tunneling bias \emph{V}$_{s}$ = + 800 mV and tunneling current \emph{I}$_{t}$ = 50 pA.
The long-range ordered CDW superstructure is evident, with the bright contrast corresponding to the CDW maxima of each David star cluster.
The statistics analysis was performed on several tens of line profiles for different images, and three typical of them are shown in Figure \textbf{S1} \textbf{f}.
The extracted periodicity of the CDW superstructure is about 1.238 $\pm$ 0.009 nm, which is larger than that of pristine 1\emph{T}-TaS$_{2}$ (1.21 nm) due to the larger atomic radius of Se.
Figure \textbf{S1} \textbf{b} indicates the Fourier transformation of the topography image \textbf{a}.
The FFT image displays strong CDW peaks with $\sqrt{13}\times\sqrt{13}$ superstructure and their higher order peaks, which rotate counterclockwise by $13.9^{\circ}$ from the primitive atomic lattice 1\emph{T}-TaS$_{2}$\cite{PhysRevB-92-085132,JPC-18-3175}.
The David star unit cells form hexagonal lattice, with the corresponding Brillouin zone denoted by the red solid lines, and one $1\times1$ unit cell marked by red dash lines.
Figure \textbf{S1} \textbf{c} displays the STM topographic image within a larger area of 60$\times$60 nm$^{2}$ at \emph{V}$_{s}$ = 800 mV and \emph{I}$_{t}$ = 50 pA.
The image includes a few different domains, with two different domain orientations by a \ang{30} rotation between them.
The domain boundaries are marked by white lines.
Figure \textbf{S1} \textbf{d} denotes the FFT image of image \textbf{c}, in which two sets of the CDW order patterns can be observed, corresponding to two different domain orientations.

\begin{figure}[H]
\centering
\includegraphics[width=12.0 cm]{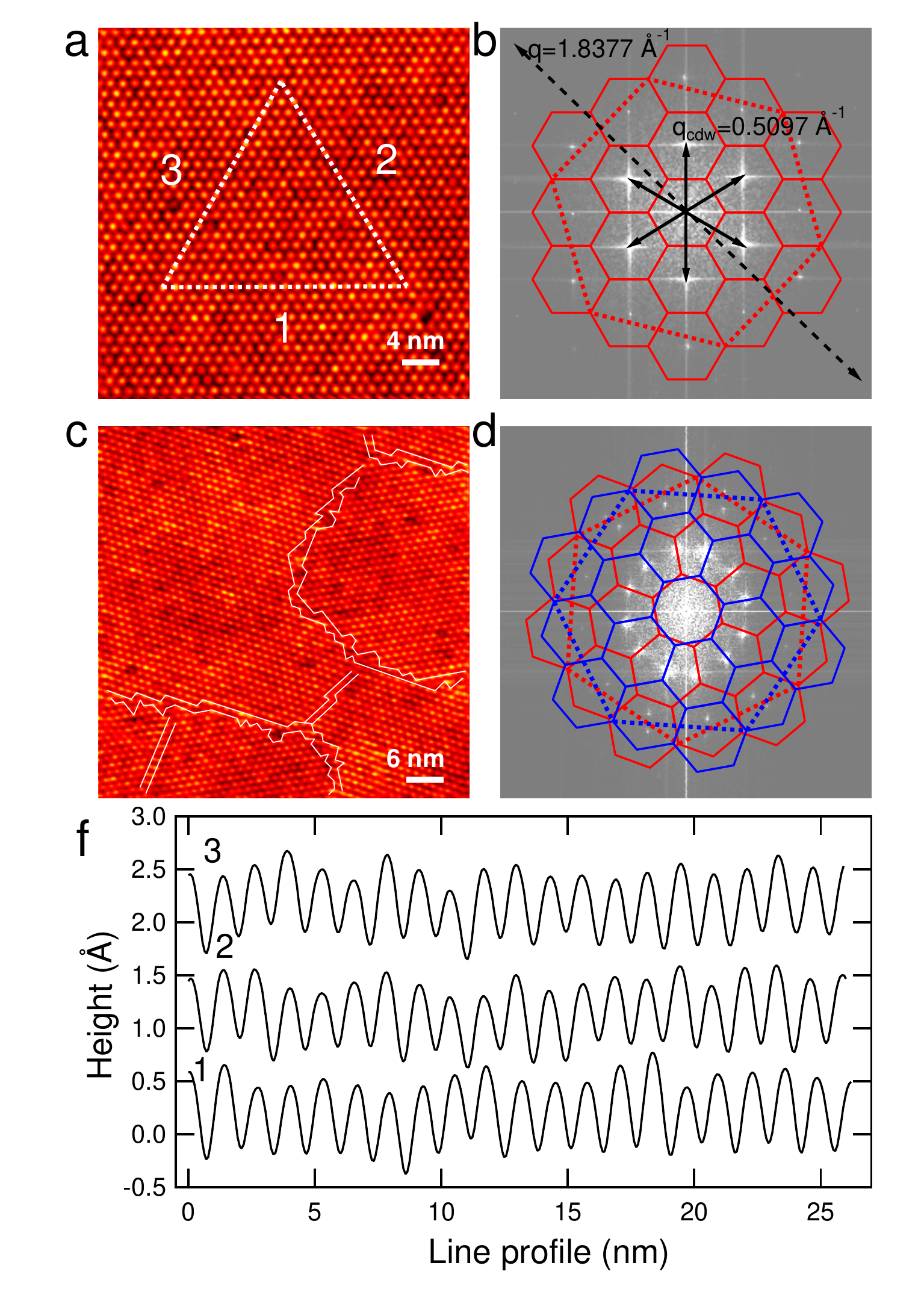}
\caption{\label{}
\textbf{a} STM topographic image taken from the 1\emph{T}-TaSSe surface over the scanning area of 40$\times$40 nm$^{2}$ within a single domain at \emph{V}$_{s}$ = 800 mV and \emph{I}$_{t}$ = 50 pA.
\textbf{b} The corresponding Fourier transformation (FFT) image.
The hexagons in solid and dashed lines correspond to the first Brillouin zone of the $\sqrt{13}\times\sqrt{13}$ R\ang{13.9} CDW structure and $1\times1$ unit cell, respectively. \textbf{c} Similar STM image taken across different domain regions with an area of 60$\times$60 nm$^{2}$ at \emph{V}$_{s}$ = 800 mV and \emph{I}$_{t}$ = 50 pA.
The domain boundaries are marked by white lines.
\textbf{d} The FFT image of \textbf{c}, in which two sets of $\sqrt{13}\times\sqrt{13}$ R\ang{13.9} CDW orders coexist, corresponding to the two rotational domains, represented by red and blue lines, respectively.
\textbf{f} The line profile of the CDW clusters as marked by the white lines in \textbf{a}, the periodicity of the CDW unit cells is 1.238 $\pm$ 0.009 nm.}

\end{figure}

\newpage

\textbf{Supplementary Note 2: Different brightness contrasts of the Se substituted 1\emph{T}-TaS$_{2}$}

Figure \textbf{S2} \textbf{a} - \textbf{p} show the series of topographic images and their corresponding FFT images.
The topographic images were acquired over a given area of 24$\times$24 nm$^{2}$ at different bias, as indicated in the images.
The hexagonal pattern of the CDW vector can be clearly observed from the FFT images, identifying that the long-ordered CDW structure is formed at low temperature for the case of 1\emph{T}-TaSSe (S:Se ratio close to 1:1), which is very similar to that of 1\emph{T}-TaS$_{2}$.
However, in contrast to the case of 1\emph{T}-TaS$_{2}$, the brightness contrast (topographic height) of each unit cell exhibits apparent variation, which was related to the different local Se concentration within the CDW unit cells \cite{PRX-7-041054}.
Figure \textbf{S3} \textbf{a} - \textbf{h} display these topographic images taken at different bias, in which the unit cells reveal different brightness contrasts.
These unitcell-by-unitcell contrasts also depend strongly on the scanning bias as shown in the figure, indicating a strong electronic effect.
The relative brightness in each image are classified into four different groups of bright, medium, dim and dark, as marked by white, green, dark blue and light blue circles in the images, respectively. As can be seen from this series of images, this brightness variation is not consistent for the images with different bias.

\begin{figure}[H]
\centering
\includegraphics[width=12.0 cm]{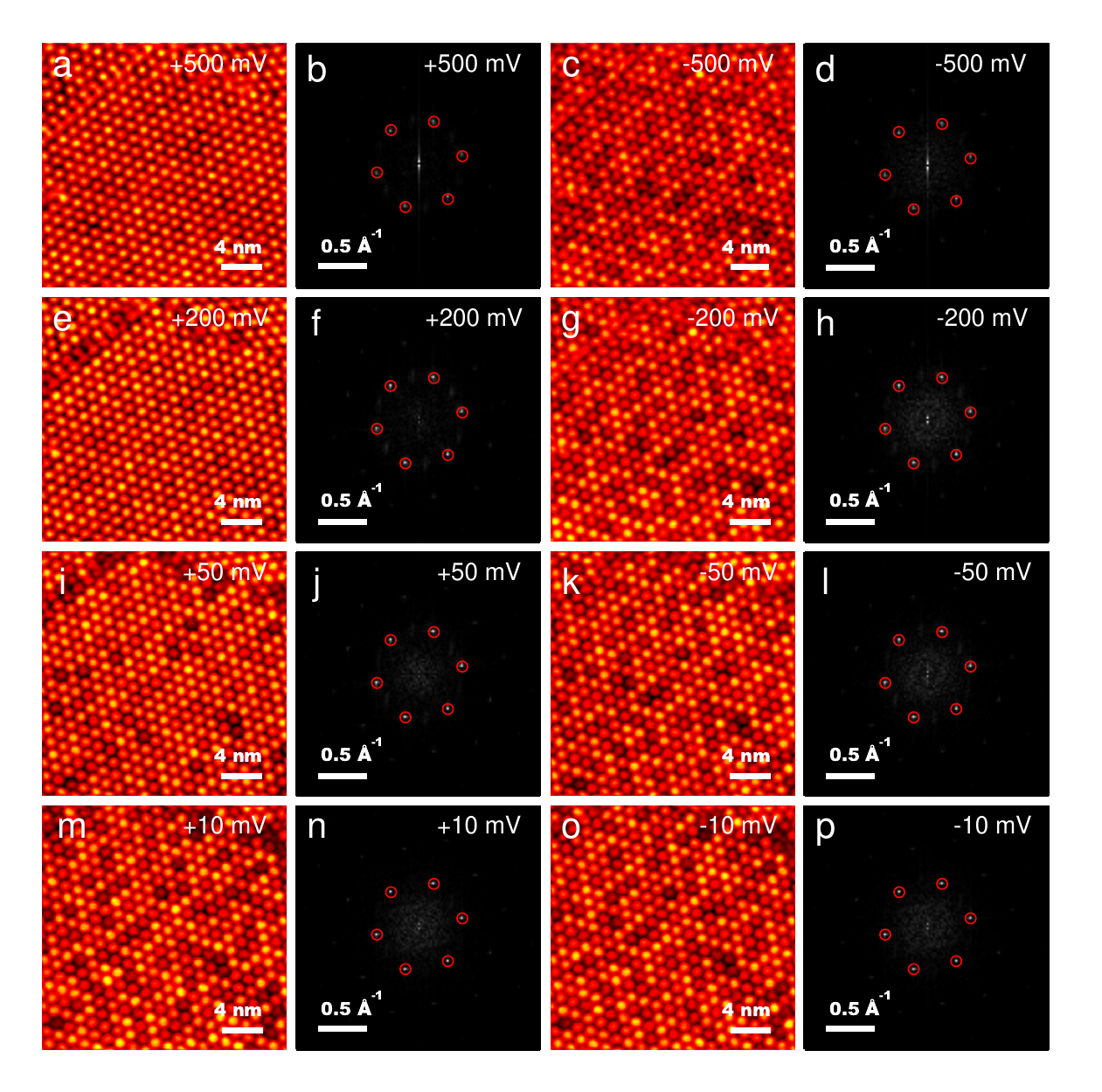}
\caption{\label{}
The STM topographic images of the 1\emph{T}-TaSSe surface taken over the scanning area of 24$\times$24 nm$^{2}$ at various sample bias, and their corresponding FFT images.
\textbf{a-b} \emph{V}$_{s}$ = + 500 mV, \emph{I}$_{t}$ = 50 pA; \textbf{c-d} \emph{V}$_{s}$ = - 500 mV, \emph{I}$_{t}$ = 50 pA; \textbf{e-f} \emph{V}$_{s}$ = + 200 mV, \emph{I}$_{t}$ = 50 pA; \textbf{g-h} \emph{V}$_{s}$ = - 200 mV, \emph{I}$_{t}$ = 50 pA; \textbf{i-j} \emph{V}$_{s}$ = + 50 mV, \emph{I}$_{t}$ = 50 pA; \textbf{k-l} \emph{V}$_{s}$ = - 50 mV, \emph{I}$_{t}$ = 50 pA; \textbf{m-n} \emph{V}$_{s}$ = + 10 mV, \emph{I}$_{t}$ = 50 pA; \textbf{o-p} \emph{V}$_{s}$ = -10 mV, \emph{I}$_{t}$ = 50 pA, as indicated in the images. }
\end{figure}

\begin{figure}[H]
\centering
\includegraphics[width=12.0 cm]{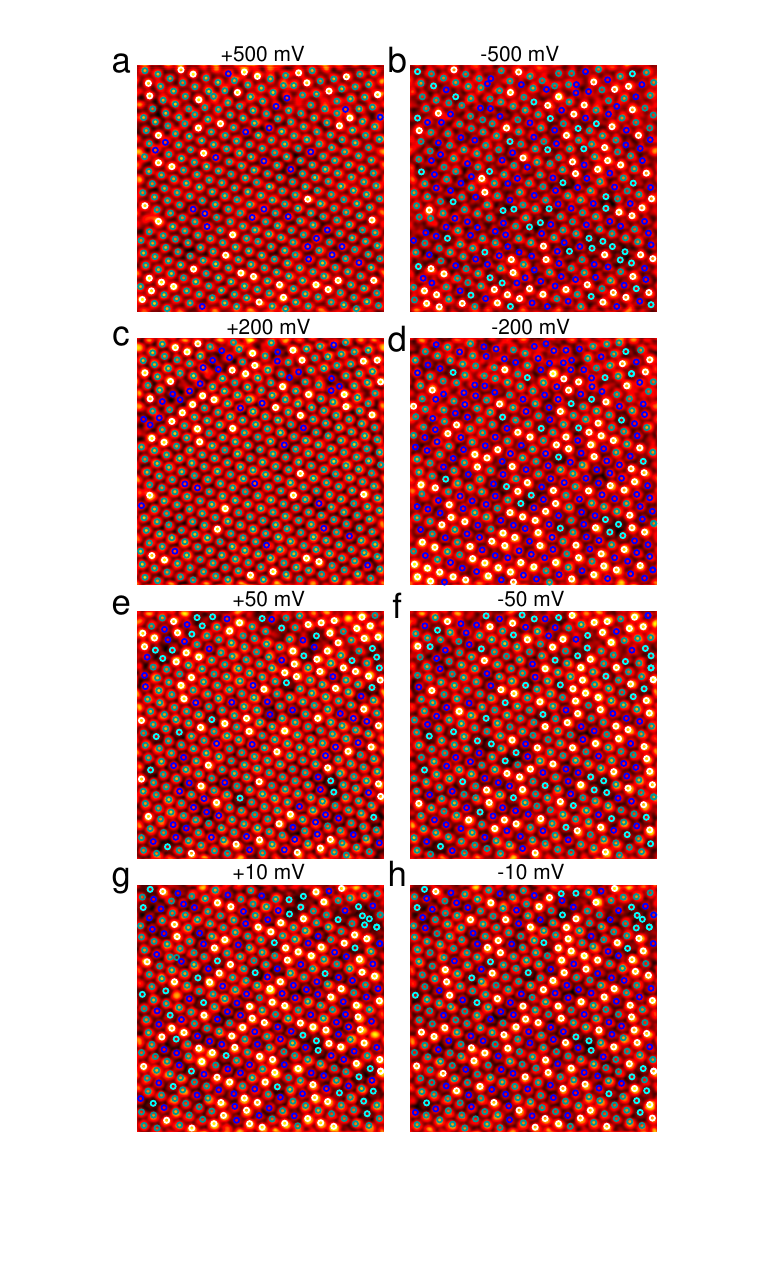}
\caption{\label{}
The STM topographic images of the 1\emph{T}-TaSSe surface with the same scanning area and scanning condition as those in Figure \textbf{S2}.
\textbf{a} \emph{V}$_{s}$ = + 500 mV, \emph{I}$_{t}$ = 50 pA; \textbf{b} \emph{V}$_{s}$ = - 500 mV, \emph{I}$_{t}$ = 50 pA; \textbf{c} \emph{V}$_{s}$ = + 200 mV, \emph{I}$_{t}$ = 50 pA; \textbf{d} \emph{V}$_{s}$ = - 200 mV, \emph{I}$_{t}$ = 50 pA; \textbf{e} \emph{V}$_{s}$ = + 50 mV, \emph{I}$_{t}$ = 50 pA; \textbf{f} \emph{V}$_{s}$ = - 50 mV, \emph{I}$_{t}$ = 50 pA; \textbf{g} \emph{V}$_{s}$ = + 10 mV, \emph{I}$_{t}$ = 50 pA; \textbf{h} \emph{V}$_{s}$ = -10 mV, \emph{I}$_{t}$ = 50 pA, as indicated in the images.
The unit cells are classified into four different groups according to the different topographic brightness. The bright, medium, dim and dark contrasts were represented by white, green, dark blue and light blue color circles, respectively. }
\end{figure}

\newpage

\textbf{Supplementary Note 3: STS spectra in the Se substituted 1\emph{T}-TaS$_{2}$}

In order to investigate the electronic structure variation over different CDW unit cells, scanning tunneling spectroscopy (STS) measurements were performed over many different unit cells.
Note that the topographic image in Figure. \textbf{S4} \textbf{a} shows the different brightness for the CDW maxima over neighboring unit cells, which indicates different Se concentration and/or configuration within each unit cell.
Figure \textbf{S4} \textbf{b}-\textbf{d} shows the typical \emph{dI/dV} spectra on CDW maxima of different unit cells, as marked in Figure. \textbf{S4} \textbf{a}.
These spectra are collected into groups according to the topographic contrasts of unit cells as classified in Figure. \textbf{S3}.
The thick (thin) solid lines correspond to the averaged (individual) \emph{dI/dV} spectra on a given group of unit cells with similar contrasts.
The individual spectra within a group show substantially larger variation between unit cells than those of pristine 1\emph{T}-TaS$_{2}$ case \cite{PhysRevB-92-085132}.
However, the averaged spectra from different contrast groups have little difference, resembling similar spectral features.
These findings indicate that the unitcell-by-unitcell variation of the spectra has a random character and is not systematically governed by the topographic contrasts.
The topographic contrasts are respected to be related to the local Se concentration/configuration.
Therefore, we can conclude that the spectra feature are not correlated with the local Se concentration (see below).

\begin{figure}[H]
\centering
\includegraphics[width=10.0 cm]{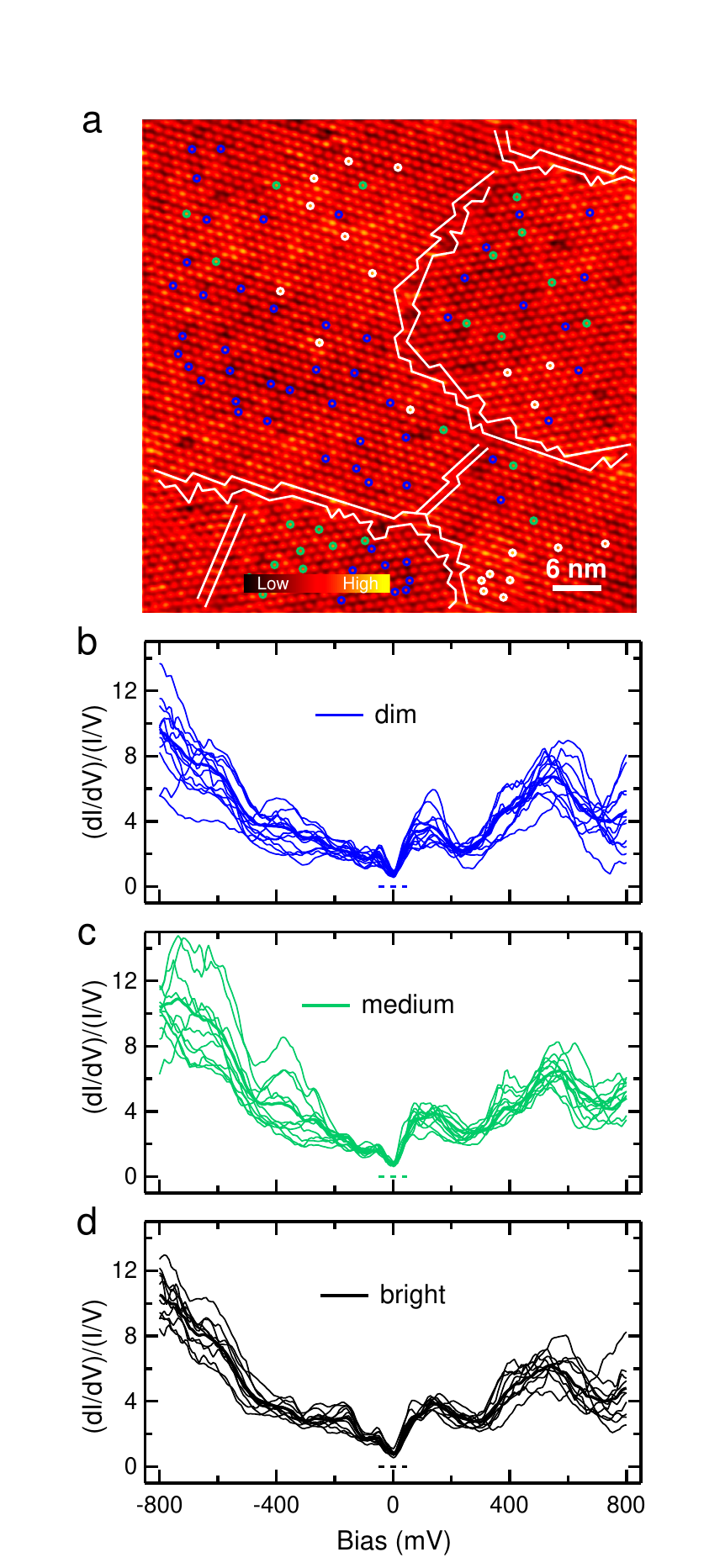}
\caption{\label{}
\textbf{a} The STM topographic image of 1\emph{T}-TaSSe taken at the area of 60 $\times$ 60 nm$^{2}$, with the scanning parameter at \emph{V}$_{s}$ = + 800 mV and \emph{I}$_{t}$ = 50 pA (the same image as Figure. 1\textbf{c} in the main text).
\textbf{b}-\textbf{d} The STS spectra of the 1\emph{T}-TaSSe.
The thick (thin) solid lines in black, green and blue color represent the averaged (individual) \emph{dI/dV} spectra taken from the CDW maxima with bright, medium and dim brightness contrasts, which are indicated by the white, green and blue circles marked in \textbf{a}, as classified in Figure. \textbf{S3}}
\end{figure}

\newpage

\textbf{Supplementary Note 4: STS spectra from the CDW maxima and minima}

More detailed spatially-resolved STS measurements were performed simultaneously with the topographic image, as shown in Figure \textbf{S5} \textbf{a} for a smaller area over about nine CDW unit cells.
Figure \textbf{S5} \textbf{b} and \textbf{c} display the \emph{dI/dV} spectra from the CDW maxima (central atoms of David star clusters) and minima regions, respectively.
The spectra from the maxima (minima) have similar main spectral features but with some intensity variation as discussed above.
Several peak features locating at around +520 and -440, +240 and -168,  + 64 and -72 meV can be observed, corresponding to the CDW gap edge, Mott and pseudogap states, respectively.

\begin{figure}[H]
\centering
\includegraphics[width=12.0 cm]{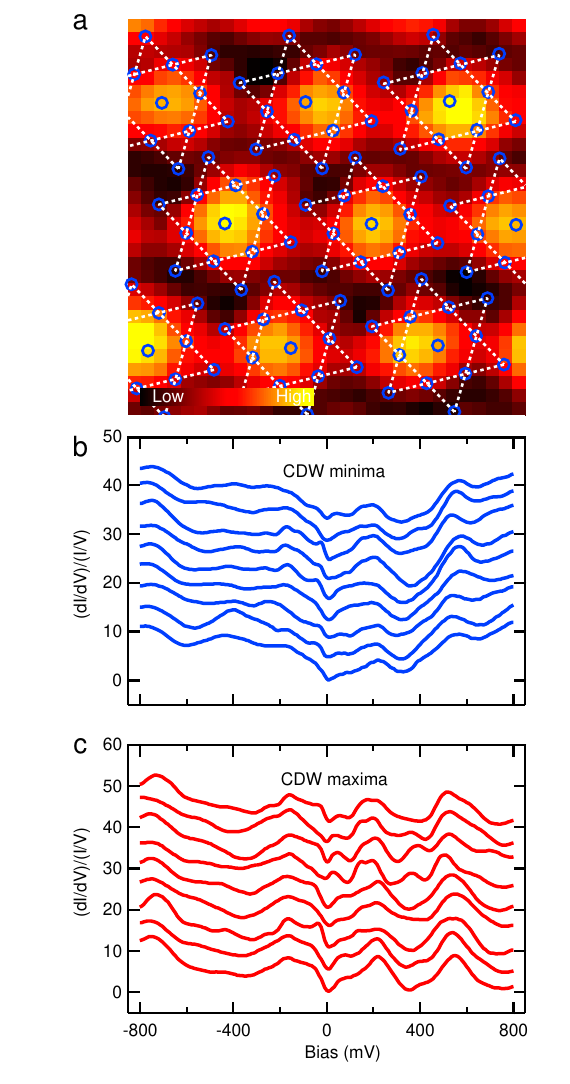}
\caption{\label{}
\textbf{a} The STM topographic image acquired with the area of 3.6$\times$3.6 nm$^{2}$ during \emph{dI/dV} measurements (\emph{V}$_{s}$ = - 800 mV and \emph{I}$_{t}$ = 900 pA). In total, \emph{dI/dV} are measured on 900 points. The star of David pattern of the CDW structure is indicated.
\textbf{b} and \textbf{c} display the normalized \emph{dI/dV} spectra from the CDW maxima and minima regions, which are denoted by red and blue colors, respectively. }
\end{figure}

\newpage

\textbf{Supplementary Note 5: DFT calculation of the electronic states of \emph{T}-TaS$_{2}$ and 1\emph{T}-TaS$_{2-x}$Se$_{x}$}

The theoretical DOS of 1$T$-TaS$_{2}$ in Figure \textbf{S6} \textbf{a} is in good agreement within 0.1 eV with the experimental $dI/dV$ (averaged) spectra.
The Mott gap with upper and lower Hubbard states at 150 and -150 meV and the CDW gap with edges at -317 and 364 meV are consistent qualitatively between the calculation and the experimental results.
The DOS of the lowest-energy 1\emph{T}-TaSSe structure (1:1 concentration of S:Se) with $\sqrt13$$\times$$\sqrt13$ periodic boundary condition in Figure \textbf{S6} \textbf{b}, however, is not consistent with the experimental $dI/dV$ spectra of disordered 1\emph{T}-TaSSe sample.
In a recent paper, the origin of the disordered phase with different brightness contrast in STM images was directly related to the Se concentration of a single David star cluster through the comparison between local $dI/dV$ spectra and theoretical DOS of the lowest-energy configurations for given Se concentrations \cite{PRX-7-041054}. This calculation is well reproduced here as shown in Figure \textbf{S6} \textbf{c}.
Considering a simple binomial distribution, the integrated probability of the Se concentration between 0.77 (10 Se atoms in a single David star) and 1.23 (16 Se atoms in a single David star) is 95 $\%$. Within this range, the DOS of lowest-energy configurations are not significantly changed in our calculations. This DOS should be similar to what is shown in Figure \textbf{S6} \textbf{b} with two well defined peaks at below and above the Fermi energy. The LHB state should be shifted down to about -300 meV, which is well localized on each CDW maxima. These characteristic features are not reproduced at all in the present experiment.
In fact, the experimentally observed spectra bear little resemblance with the calculated DOS of any given concentration. These discrepancy clearly indicates that the electronic structure of the disordered 1\emph{T}-TaSSe can not be described by the DFT+U calculations for the periodic structure of a given concentration and a given S-Se configuration, due to the electron correlation and the disorder effect.
As shown below, the atomic configurations (positions of substituted Se atoms) are as important as the Se concentration or configuration for the electronic structure of 1\emph{T}-TaS$_{2-x}$Se$_{x}$.

\begin{figure} [H]
\includegraphics[width=16.0 cm]{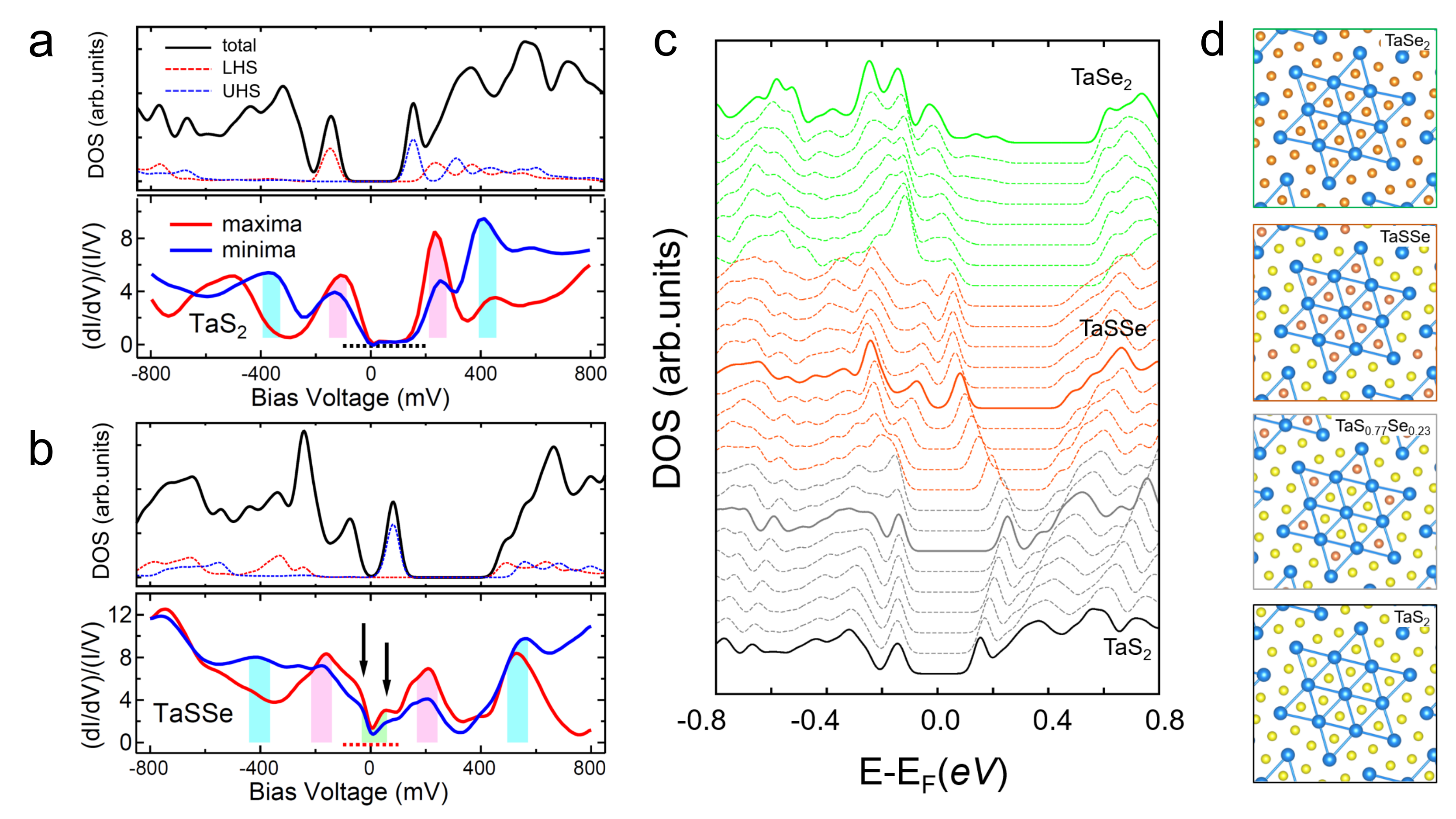}
\caption{\label{}
Theoretical density of states (DOS) and experimental $dI/dV$ spectra of 1$T$-TaS$_{2}$ and 1$T$-TaS$_{2-x}$Se$_{x}$.
\textbf{a} Theoretical DOS (upper panel) and experimental $dI/dV$ spectra (bottom panel) of 1$T$-TaS$_{2}$.
Red and blue lines are the spin-polarized local DOS ($\times$3) at the center Ta atom of David star indicating the lower and upper Hubbard states, respectively.
Black solid line denotes the total DOS.
\textbf{b} Theoretical DOS (upper panel) of the lowest-energy TaSSe structure with the $\sqrt{13}\times\sqrt{13}$ unit cell.
The bottom panel is the experimental $dI/dV$ spectra of 1$T$-TaSSe.
\textbf{c} Total DOS of the lowest-energy structures of  1$T$-TaS$_{2-x}$Se$_{x}$.
\textbf{d} Selected atomic configurations.
}
\end{figure}

\newpage

\textbf{Supplementary Note 6: High-symmetry atomic configurations and calculated DOS}

We show that the electronic structure of 1\emph{T}-TaSSe depends substantially on the S-Se atomic configuration for a given Se concentration, too. Figure \textbf{S7} \textbf{a} shows the high-symmetry atomic configurations among various possible ones of the 1\emph{T}-TaSSe (1:1 concentration of S:Se) structure with the $\sqrt13$$\times$$\sqrt13$ periodic boundary condition.
As shown in Figure \textbf{S7} \textbf{b}, the upper and lower Hubbard states and the CDW gap edge states are highly affected by the Se configurations and the difference between different configurations are as large as that for different concentrations shown in Figure \textbf{S6}.
This part was not considered properly in the previous work \cite{PRX-7-041054}.
Averaged DOS of these atomic configurations in Figure \textbf{S7} \textbf{c} is comparable to the experimental (averaged) $dI/dV$ spectra in a much better agreement than the averaged one over different Se concentrations (for a binomial distribution) shown in Figure \textbf{S6} \textbf{c}, while they still miss the most important pseudogap feature.
We may suggest that the major source of the disorder can be the configurational disorder instead of the Se concentration.
Further investigation is requested in this aspect.

\begin{figure} [H]
\includegraphics[width=16.0 cm]{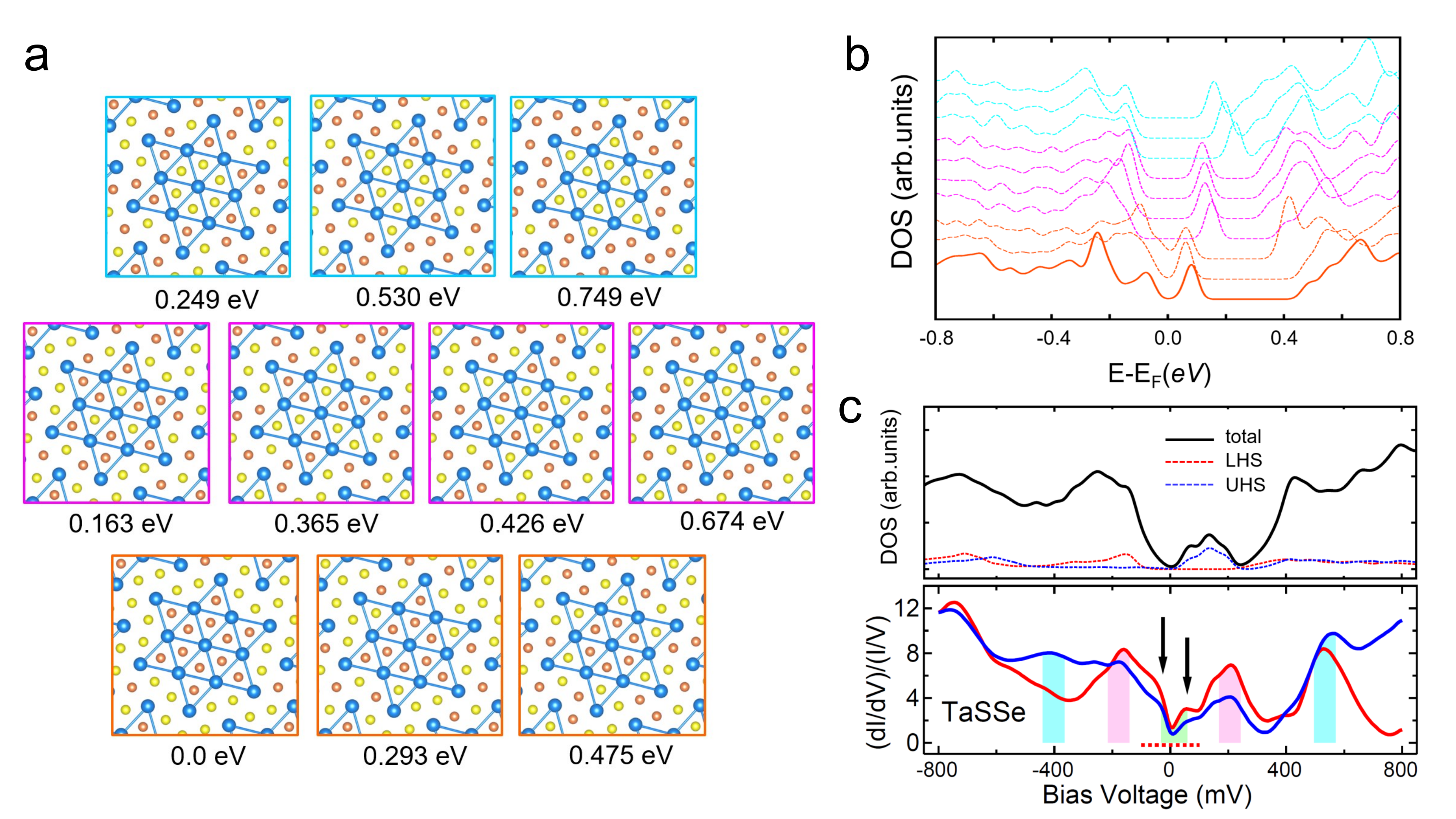}
\caption{\label{}
\textbf{a} High-symmetry atomic configurations of TaSSe with a relative energy.
\textbf{b} Total DOS of the TaSSe configurations in \textbf{a}.
\textbf{c} Averaged total DOS of the TaSSe configurations with the experimental $dI/dV$ spectra of TaSSe.
}
\end{figure}

\newpage

\textbf{Supplementary Note 7: Spatial fluctuation of the electronic states}

Figure \textbf{S8} \textbf{a}-\textbf{f} display the LDOS maps over a scanning area of 24$\times$24 nm$^{2}$ at different scanning bias.
The CDW superstructure patterns are clearly seen from all the LDOS maps, but the electronic states have obvious fluctuation between CDW unit cells, and this variation is randomly disordered.
We estimate the spatial fluctuation of the electronic states along the dashed lines in the images. The amplitude of this fluctuations is about 20-40 \% of the average LDOS values, as shown in Figure \textbf{S8} \textbf{g}.

\begin{figure}[H]
\centering
\includegraphics[width=12.0 cm]{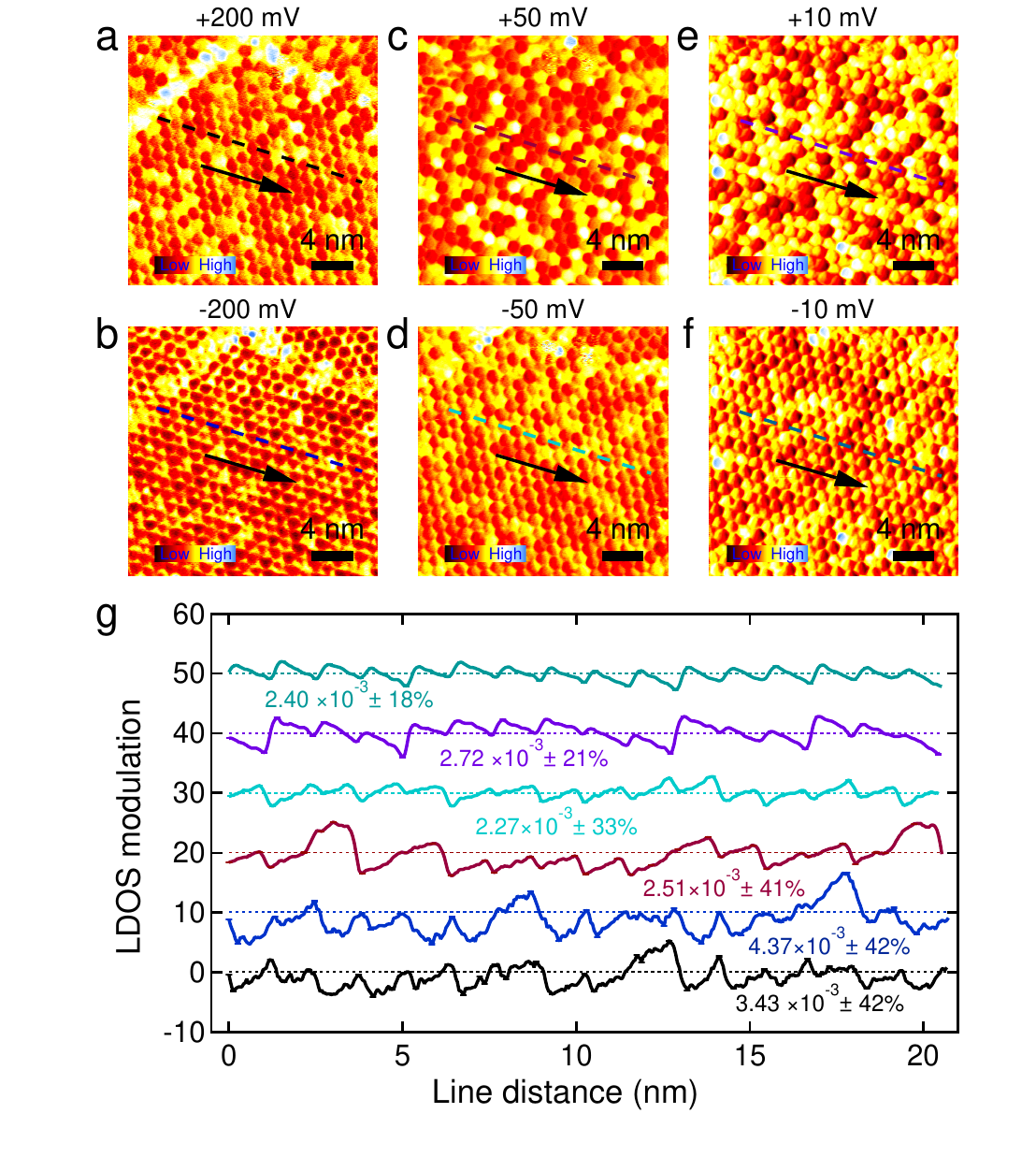}
\caption{\label{} The LDOS maps were taken over the scanning area of 24$\times$24 nm$^{2}$ at various sample bias of \textbf{a} \emph{V}$_{s}$ = + 200 mV, \emph{I}$_{t}$ = 50 pA; \textbf{b} \emph{V}$_{s}$ = - 200 mV, \emph{I}$_{t}$ = 50 pA; \textbf{c} \emph{V}$_{s}$ = + 50 mV, \emph{I}$_{t}$ = 50 pA; \textbf{d} \emph{V}$_{s}$ = - 50 mV, \emph{I}$_{t}$ = 50 pA; \textbf{e} \emph{V}$_{s}$ = + 10 mV, \emph{I}$_{t}$ = 50 pA; \textbf{f} \emph{V}$_{s}$ = -10 mV, \emph{I}$_{t}$ = 50 pA. \textbf{g} shows the spatial fluctuation of the electronic states along the dashed lines marked in \textbf{a}-\textbf{f}.}
\end{figure}

\newpage

\textbf{Supplementary Note 8: Multifractal spectrum \emph{f}($\alpha$)}

Based on the theoretical predictions, the critical states are expected to represent the multifractal spatial structure, which is related to the scale-invariant nature of the wave function.
The multifractal structure is usually described by analysis of the self-similarity through the singularity spectrum \emph{f}($\alpha$).
Physically, \emph{f}($\alpha$) describes all the fractal dimensions that share a common exponent $\alpha$ in a spatial pattern, where the eigenfunction intensity satisfy $ \left|\Psi^{2} (\vec{r})\right| \sim \emph{L}^{-\alpha}  $\cite{RMP-80-1355,Science-327-665,PRL-62-1327}.
A variety of techniques have been developed to calculate \emph{f}($\alpha$).
Here, the multifractal spectra is calculated by following the method developed in Ref. \cite{PRL-62-1327}.
The spatial probability distribution of the wave function
\begin{equation}
P_{q}=\int \ \vert\Psi(r)\vert ^{2q}\, d ^d r \varpropto L^{-\tau{_{q}}}
\end{equation}
The multifractal spectrum can be achieved by the Legendre transform

\begin{equation}
f(q)=\alpha q-\tau(q)
\end{equation}

\begin{equation}
\alpha=\frac{d\tau(q)}{dq}
\end{equation}
According to the method proposed by Chahabra and Jensen, the multifractal spectra \emph{f}($\alpha$) can be calculated as follows

\begin{equation}
f(q)=\lim_{L\to 0} \frac{\sum \limits_{i} \mu_{i}(q,L)\log[\mu_{i}(q,L)]}{\log L}
\end{equation}

\begin{equation}
\alpha(q)=\lim_{L\to 0} \frac{\sum\limits_{i} \mu_{i}(q,L)\log[P_{i}(L)]}{\log L}
\end{equation}

Figure 9 \textbf{a-c} shows the \emph{f}($\alpha$) calculated by using the box-counting technique \cite{Science-327-665,PRL-62-1327} for the LDOS maps at different energies.
For all the energies, the \emph{f}($\alpha$) spectra exhibit finite width and parabolic shape, with the maximum values locating at the position of $ \alpha_{0} = d + \epsilon $, where $ d = 2.0 $ and $ \epsilon \ll 1$.
The multifractal spectra exhibit parabolic shape together with the maximum position at $ d + \epsilon $ with $ \epsilon \ll 1$, demonstrating that the LDOS exhibits the weak multifractal behavior \cite{RMP-80-1355}. The multifractality would be strong in non-interaction systems but be suppressed in interacting systems as the present case.
The weak multifractal behavior together with the Gaussian distribution identify the disordered extended states as expected in the two-dimensional electronic systems.

\begin{figure}[H]
\centering
\includegraphics[width=12.0 cm]{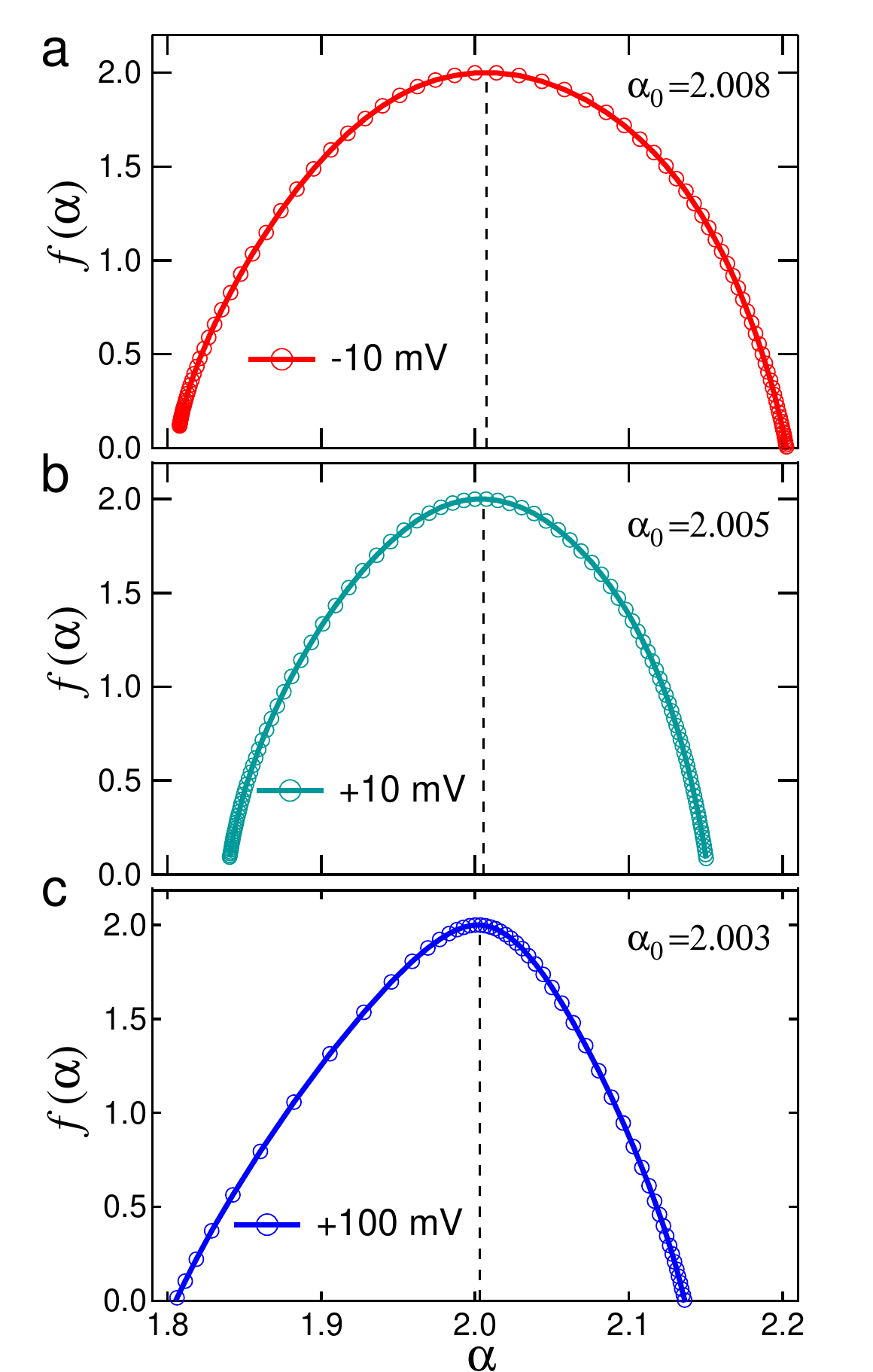}
\caption{\label{} The multifractal spectra calculated from the LDOS maps at -10 , +10 and +100 mV as indicated by red, blue and green color lines. All the multifractal spectra \emph{f}($\alpha$) exhibit parabolic shape and have the maximum values at $ \alpha_{0} = d + \epsilon $, where $ d = 2.0 $ and $ \epsilon = 0.008, 0.005, 0.003$ at different energies.}
\end{figure}